\documentclass[1p]{elsarticle}
\usepackage{hyperref}
\usepackage{float}
\usepackage{tabulary}
\usepackage{csquotes}
\usepackage{booktabs}
\usepackage{multirow}
\usepackage{hyperref}
\usepackage{amsmath}
\usepackage[table]{xcolor}
\usepackage[ruled]{algorithm2e}

\definecolor{Gray}{gray}{0.9}

\journal{Journal of Systems and Software}

\begin{document}

\begin{frontmatter}

\title{Test Prioritization in Continuous Integration Environments}

\address[label1]{Empirical Software Engineering in Software, Systems and Services (M3S), Faculty of Information Technology and Electrical Engineering (ITEE), University of Oulu, Oulu, Finland.}
\cortext[cor1]{Corresponding author. Tel.: +358 46 920 47 11.\\
E-mail address: alireza.haghighatkhah@oulu.fi (Alireza Haghighatkhah).}

\author[label1]{Alireza Haghighatkhah\corref{cor1}}
\author[label1]{Mika M\"{a}ntyl\"{a}}
\author[label1]{Markku Oivo}
\author[label1]{Pasi Kuvaja}

%% ********************************************************************************************
%% ABSTRACT
%% ********************************************************************************************
\begin{abstract}
Two heuristics namely diversity-based (DBTP) and history-based test prioritization (HBTP) have been separately proposed in the literature. Yet, their combination has not been widely studied in continuous integration (CI) environments. The objective of this study is to catch regression faults earlier, allowing developers to integrate and verify their changes more frequently and continuously. To achieve this, we investigated six open-source projects, each of which included several builds over a large time period. Findings indicate that previous failure knowledge seems to have strong predictive power in CI environments and can be used to effectively prioritize tests. HBTP does not necessarily need to have large data, and its effectiveness improves to a certain degree with larger history interval. DBTP can be used effectively during the early stages, when no historical data is available, and also combined with HBTP to improve its effectiveness. Among the investigated techniques, we found that history-based diversity using NCD Multiset is superior in terms of effectiveness but comes with relatively higher overhead in terms of method execution time. Test prioritization in CI environments can be effectively performed with negligible investment using previous failure knowledge, and its effectiveness can be further improved by considering dissimilarities among the tests.
\end{abstract}

\begin{keyword}
Test case prioritization\sep regression testing\sep continuous integration\sep
build history\sep test diversity
\end{keyword}

\end{frontmatter}

%%\linenumbers

%% ********************************************************************************************
%% CHAPTER 1. INTRODUCTION
%% ********************************************************************************************
\section{Introduction}

The software industry is moving toward an agile, continuous delivery paradigm in which changes to software are released more frequently and considerably faster than before \cite{rodriguez2017continuous,mantyla2015rapid}. To make the rapid evolution of software cost-effective and reliable, the industry has adopted continuous integration (CI) \cite{duvall2007continuous}. CI aims to prevent the integration problem known as \enquote{integration hell} and automate the build process and verification of changes. Each integration cycle is called a \textit{build}. The build comprises a set of automated activities and is followed by regression testing (RT). In a nutshell, RT aims to ensure that recent changes to the system have not impacted negatively on any previously verified functionality. RT is widely used in practice; it is common to have a dedicated regression test suite that is often run in its entirety \cite{engstrom2010qualitative}. 

A test suite for enterprise-sized applications often includes thousands of test cases, the execution of which requires several hours or even days. For instance, the JOnAS Java EE middleware comprises 2,689 test cases \cite{kessis2005experiences}. Applying them all to its 16 configurations results in running 43,024 test cases. Furthermore, the cost of RT increases over the time with the increase in system size. Memon et. al \cite{memon2017taming} observed linear growth in both code submission rate and size of regression test suite, so incurring significant expenses to keep the RT running. The software engineering literature has proposed many techniques to improve RT processes. Test suite minimization (TSM) \cite{harrold1993methodology} aims to eliminate test cases from a test suite with a specific objective, i.e., removing obsolete or redundant test cases. Several experiments have been reported in the literature with differing conclusions regarding the impact of TSM on the fault detection capability of a test suite, e.g., \cite{rothermel1998empirical,wong1999test,heimdahl2007effect}. However, the common understanding is that TSM may compromise such capability. Test case prioritization (TCP) \cite{rothermel2001prioritizing}, on the other hand, is concerned with the ideal ordering of test cases to maximize desirable properties (i.e., early fault detection). From the perspective of fault detection, TCP seems to be a safe approach because it does not eliminate test cases and simply permutes them within the test suite. 

The intersection of CI and RT poses great challenges for the software development industry. The testing budget is often limited and RT needs to make the most of sometimes limited resources. For RT improvement techniques to be useful and easy for the industry to adopt, they must consider contextual factors related to enterprise-level testing environments \cite{do2016chapter}. Several TCP techniques have been proposed that can be applied in a CI environment\cite{kim2002history,elbaum2014techniques,marijan2013test,strandberg2016experience,srikanth2016test,hemmati2017prioritizing,spieker2017reinforcement}. Even though differences exist among the proposed techniques, they all share a common assumption: tests which previously failed are much more likely to fail again. Empirical studies to support such a heuristic are emerging in the literature. For instance, in 2017, Hemmati et al. \cite{hemmati2017prioritizing} investigated the effectiveness of three black-box TCP techniques on Firefox before and after the transition to rapid releases. The authors concluded that history-based test prioritization (HBTP) is far more effective than other comparable techniques in rapid release, although this does not hold in traditional development environments. TCP using previous failure knowledge, however, comes with its own limitations. For instance, not all regression faults can be captured effectively, if there has been no previous failure, e.g., newly added test cases or those which have not previously revealed any failure. To increase the likelihood of capturing faults, one potential strategy might be to spread the testing budget evenly across different parts of the system by performing a diverse set of test cases \cite{ledru2012prioritizing}. The underlying assumption is that similar test cases will likely exercise the same part of the system and detect the same fault; thus, a diverse set of test cases should be performed to detect a greater number of faults \cite{jiang2009adaptive}. Diversity-based TCP requires minimal information, since the only required information is already encoded in the test suite \cite{ledru2012prioritizing}.

In this work, we classify regression faults according to their past verdicts and study the extent to which they can be captured using previous failure knowledge. To effectively deploy HBTP, one might assume that a large amount of historical data is required. More specifically, history interval size (e.g., the number of previous verdicts used) is often not reported in previous works, and its impact on the effectiveness of HBTP has not been studied. In this work, we perform HBTP using differing numbers of previous verdicts and investigate whether its effectiveness changes by varying the size of the history interval. To increase the likelihood of capturing faults, one potential strategy is to assign a higher priority to those test cases which are most different compared to those already prioritized. Even though diversity-based TCP has been proposed and used previously, it has not been widely studied in combination with HBTP and in a CI environment. Hemmati et al. \cite{hemmati2017prioritizing} investigated a history-based diversity approach in Firefox project. They used previous failure knowledge in combination with a single similarity metric, Manhattan distance, using the English texts of manual test cases. In contrast to their study, we investigated history-based diversity using three different similarity metrics (Manhattan, normalized compression distance (NCD), and NCD Multiset) with different history interval sizes, over a large number of automated builds extracted from six open-source projects. In this study, we examine whether the effectiveness of HBTP is impacted when it is combined with diversity-based TCP. For the history-based diversity technique to be applicable in a CI environment, both effectiveness and performance are critical. Thus, in addition to effectiveness, we investigated the method execution time within and across the studied projects.
 
The main objective of our study is to catch regression faults earlier, allowing developers to integrate and verify their changes more frequently and continuously. To achieve this, we investigated six open-source software projects; each project included several builds over a long period of time. Findings from our study indicate that:

\begin{itemize}
\item Only a small proportion of tests has ever failed with our subjects (less than 11\% in four projects, and 3\textendash 52\% overall). This indeed raises the importance of TCP in CI environments in which RT is performed more frequently and continuously. Furthermore, the majority of regression faults (57\textendash 97\%) among all investigated projects can be captured solely by using previous failure knowledge. This implies that previous failure knowledge seems to have strong predictive power in CI environments and can be used to effectively prioritize tests.

\item HBTP does not necessarily require a large amount of historical data, and its effectiveness improves to a certain degree with a larger history interval. Even with the last verdict (current$-1$), improvement (Vargha\textendash Delaney A measure: 0.53\textendash 0.82) in terms of average percentage of faults detected was observed in all studied projects in comparison to random ordering.

\item Diversity-based TCP can be used effectively during the early stages, in which no historical data is available (Vargha\textendash Delaney A measure: 0.68\textendash 0.91, improvement observed using NCD Multiset), or combined with HBTP to improve its effectiveness (Vargha\textendash Delaney A measure: 0.51\textendash 0.73 using NCD Multiset).

\item Among the investigated history-based diversity techniques, i.e., pairwise Manhattan, pairwise NCD, and NCD Multiset, we found that the latter is superior in terms of effectiveness but comes with relatively higher overhead in terms of method execution time. 
\end{itemize}

From academic point of view, we provide empirical evidence in support of two previously proposed heuristics namely history-based and diversity-based TCP in CI environments. From the perspective of practitioners, our findings indicate that HBTP can be employed with negligible investment and its effectiveness can be further improved by considering distances (dissimilarities) among the tests. The rest of this paper is organized as follows: the background and related work is presented in Section 2. Section 3 describes the research methodology, while Section 4 presents the results of our study. Section 5 discusses the findings and their implications, including a discussion of the validity of this research and concluding remarks are given in Section 6.

%% ********************************************************************************************
%% CHAPTER 2. RELATED WORK
%% ********************************************************************************************
\section{Background and Related Work}

In this section, we briefly review various RT improvement techniques, explain the related work, and highlight our own contribution in section \ref{section_researchgap}.

\subsection{Background}
\begin{figure}[H]
\centering
\includegraphics[width=1\columnwidth]{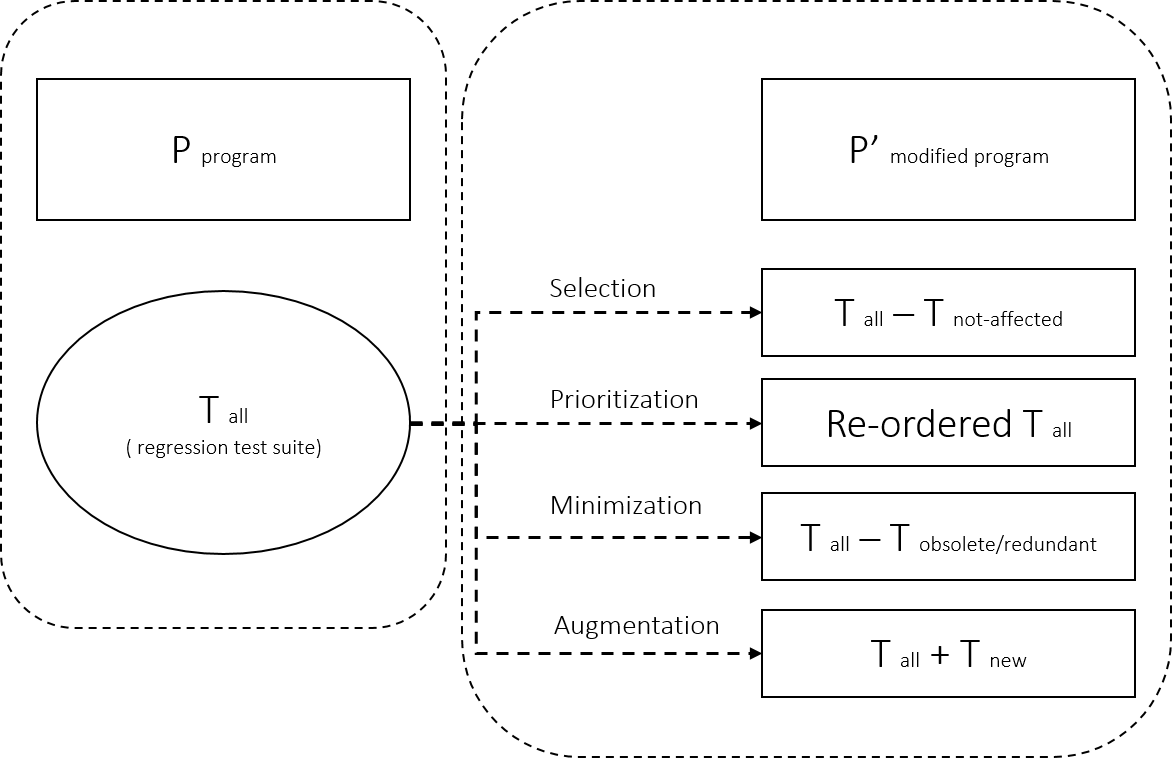}
\caption{General model of RT techniques (adapted from \cite{do2016chapter})} \label{fig:RT_Background}
\end{figure}

A number of RT techniques and tools have been developed and proposed as approaches to reducing expenses and improving processes. Yoo and Harman\textsc{\char13}s \cite{yoo2012regression} comprehensive survey reviewed RT techniques originally introduced by Rothermel and Harrold \cite{harrold1993methodology,rothermel1996analyzing,rothermel1998empirical,rothermel1999test}. Figure \ref{fig:RT_Background} depicts a general model of RT techniques. 

Let $P$ be a program, $P'$ be a modified version of the program, and $T$ be a test suite developed for $P$. In the transition from $P$ to $P'$, a previously verified behavior of $P$ may have become faulty in $P'$. RT seeks to validate $P'$ to ensure that recent changes to the system have not impacted negatively on any previously verified functionalities. During RT, several techniques may be employed in practice. TSM seeks to identify and permanently eliminate obsolete or redundant test cases from the test suite. Regression test selection (RTS) aims to select only the subset of test cases affected by the recent changes. TCP is concerned with the ideal ordering of test cases to maximize desirable properties (i.e., early fault detection), while test suite augmentation aims to identify newly added source code and to generate new test cases accordingly.

There is a large body of research on RT, with a great deal of pioneering work published in the 1990s \cite{harrold1993methodology,rothermel1996analyzing,rothermel1998empirical,rothermel1999test}. In 2016, Garousi and M\"{a}ntyl\"{a} \cite{garousi2016systematic} conducted a tertiary study of systematic literature reviews of software testing and identified 11 published secondary studies on various aspects of RT. The most recent systematic literature review on TCP techniques was provided by Khatibsyarbini et al. \cite{khatibsyarbini2017test} in 2017. The authors classified existing TCP approaches into nine categories. They concluded that TCP approaches are still broadly open for improvement; each approach has specific potential values, advantages, and limitations. Despite the large body of knowledge on RT and the many academic advancements, there is relatively little evidence regarding the practical application of RT techniques in industrial settings \cite{orso2014software}. The industry practice seems to be based mostly on experience rather than on any systematic approach to RT \cite{engstrom2010qualitative}. To ensure RT techniques are useful and easily adopted, they must consider contextual factors related to enterprise testing environments \cite{do2016chapter}.

\subsection{Related Work}

\subsubsection{Test Prioritization in CI Environments}

Several TCP techniques have been proposed that can be applied in CI environments \cite{kim2002history,elbaum2014techniques,marijan2013test,strandberg2016experience,srikanth2016test,hemmati2017prioritizing,spieker2017reinforcement}. Even though differences exist among the proposed techniques, they all share a common heuristic and rely on previous failure knowledge. Elbaum et al. \cite{elbaum2014techniques} argued that traditional RT techniques tend to rely on code instrumentation and are applicable only to discrete, complete sets of test cases. However, RT in CI environments is performed more frequently and continuously. Thus, approaches that require exhaustive analysis are overly expensive and inefficient due to the high frequency of changes; this makes the data gathered by such approaches imprecise and obsolete. Elbaum et al. \cite{elbaum2014techniques} conducted an empirical study on a large data set obtained from Google and presented novel RTS and TCP techniques. The proposed approaches are based on the notion of time windows to track how recently test suites were executed and revealed failures. This information was utilized to select test suites to be executed during pre-submit testing and to prioritize test suites that must be performed during post-submit testing.

Marijan et al. \cite{marijan2013test} proposed a TCP technique which relies on previous failure knowledge, test execution time, and domain-specific heuristics to compute the test priority using a weighted function. Strandberg et al. \cite{strandberg2016experience} presented an experience report and proposed an automated tool which aimed to combine priorities of multiple factors associated with test cases. Spieker et al. \cite{spieker2017reinforcement} conducted industrial case studies and proposed an approach that used reinforcement learning to select and prioritize test cases according to their duration, previous last execution, and failure history. Srikanth et al. \cite{srikanth2016test} conducted an empirical study on TCP for the build acceptance test process of a large enterprise software-as-a-service application. Findings indicated that ordering build acceptance tests can significantly impact efficiency of testing and that the use of historical data is a good heuristic for test prioritization.

Hemmati et al. \cite{hemmati2017prioritizing} investigated the effectiveness of three black-box TCP techniques on Firefox before and after the transition to rapid releases. The authors concluded that HBTP is far more effective than other comparable techniques in rapid releases, although the same conclusion does not hold in traditional development environments. Therefore, the tests that failed in previous releases have a much higher probability of failing again in the current release. This is perhaps due to the recency of historical knowledge, which explains its effectiveness in a rapid-release environment, rather than other changes in the development process \cite{hemmati2017prioritizing}. There is also a number of studies that are particularly focused on HBTP. For instance, Kim and Porter \cite{kim2002history} proposed a TCP technique that relies on test execution history, its fault detection, and the program entities it covers. Following this study, others \cite{park2008historical,fazlalizadeh2009prioritizing,khalilian2012improved} investigated HBTP and proposed various models to compute the priority of test cases using previous failure knowledge and other complementary information.

\subsubsection{Diversity-based Test Prioritization}

Coverage-based TCP has been extensively studied in the literature (see the latest systematic literature review on TCP \cite{khatibsyarbini2017test}). One common heuristic is to assign a higher rank to the test cases that cover a part of the system that has not been examined earlier by other test cases \cite{rothermel2001prioritizing}. To increase the likelihood of capturing faults, one potential strategy is to spread the testing budget evenly across different parts of the system by performing a diverse set of test cases. Diversity-based TCP has been previously proposed in the literature. For instance, Leon and Podgurski \cite{leon2003comparison}, proposed a TCP technique using automatic cluster analysis to partition the pool of test cases based on how their execution profiles are distributed in the profile space. This technique was later extended by Yoo et al. \cite{yoo2009clustering} to incorporate the domain expert knowledge. More recently, Ledru et al. \cite{ledru2012prioritizing} employed string metrics to measure the similarities among test cases. 

The underlying hypothesis is that similar test cases will likely exercise the same part of the system and detect the same fault; thus, a diverse set of test cases must be performed to detect more faults \cite{jiang2009adaptive}. This hypothesis has been further investigated by other researchers, e.g.,\cite{cartaxo2011use,hemmati2010reducing,hemmati2011empirical,hemmati2013achieving,hemmati2017prioritizing,ledru2012prioritizing,arafeen2013test,thomas2014static}. The implication for TCP is that higher priority must be assigned to those test cases that are most different from those already prioritized. Diversity-based TCP can be implemented using different methods and on different levels, e.g., source-code behind test cases \cite{ledru2012prioritizing}, method calls \cite{hemmati2013achieving}, topic models extracted from test cases \cite{thomas2014static}, or English texts of manual test cases \cite{hemmati2017prioritizing}. 

To achieve diversity-based TCP, the dissimilarity among test cases must be calculated using a particular method; this information must then be leveraged to prioritize test cases. There are several similarity metrics proposed in the literature and used in diverse areas, e.g., classification problems, plagiarism detection, and image and DNA analysis. It is commonly understood that similarity metrics have different characteristics and are typically specific to a certain type of data. Ledru et al. \cite{ledru2012prioritizing} conducted a comprehensive experiment on \enquote{Siemens Test Suite} and evaluated four classical string metrics for the purpose of TCP, including Cartesian, Levenshtein, Hamming, and Manhattan distance. Their findings indicated that TCP using string metrics is more effective than a randomly ordered test suite, and Manhattan distance yields better results than the other investigated metrics. To calculate the distance between a test case $t$ and set of test cases $T'$, Ledru et al. proposed the following function which uses distance measure $d$:

$$AllDistances(t, T', d) = min \{d(t, t_i)| t_i \in T', t_i \neq t\} $$

Ledru et al. used the min operation because an empirical study by Jiang et al. \cite{jiang2009adaptive} showed that maximize-minimum is more efficient in comparison to maximize-average and maximize-maximum. Ledru et al. also proposed a greedy algorithm that iteratively picks a test case having maximum distance (i.e., is most dissimilar) to the set of already prioritized test cases. NCD is another similarity metric which is universal and can be applied to any strings of data, regardless of data type investigated \cite{li2004similarity,cilibrasi2005clustering}. NCD has been extensively used in a wide range of application areas (see the many references in Google Scholar to \cite{li2004similarity,cilibrasi2005clustering}). NCD is given as follows, where $C$ is a function that calculates the approximate Kolmogorov complexity and returns the length of the input string after its compression, using a chosen compression program:

$$NCD(x,y) = \frac{ C(xy) - min \{C(x), C(y)\}} { max \{C(x), C(y)\} } $$

In 2015, Cohen and Vit\'anyi \cite{cohen2015normalized} extended the application of NCD for multisets (a particular type of set which allows multiple instances). NCD Multiset provides similarity measurement at the level of entire sets of elements rather than between pairs. Feldt et al. \cite{feldt2016test} performed the first study in software engineering literature that used NCD Multiset. They conducted an experiment, the results of which show that test selection using NCD Multiset leads to higher structural and fault coverage in comparison to random selection. NCD Multiset has also been used recently in TCP literature \cite{henard2016comparing} and the results seem to be promising.

\label{section_researchgap}
\subsubsection{Research Gap}
In comparison to previous studies, we make three contributions. First, we thoroughly investigate the impact of the history interval size on the effectiveness of HBTP. It might be assumed that a large set of historical data is required to effectively deploy HBTP in practice. Specifically, history interval size (i.e., the number of previous verdicts used) is often not reported in previous works and its impact on the effectiveness of HBTP has not been studied. In this work, we perform HBTP using differing numbers of previous verdicts and investigate whether its effectiveness changes by varying the size of the history interval. Furthermore, we classify regression faults according to their past verdicts and study the extent to which they can be captured using previous failures knowledge.
 
Second, to improve the effectiveness of HBTP, we combine previous failure knowledge with diversity-based TCP. Prior works have used previous failure knowledge with code-coverage \cite{yoo2011faster,epitropakis2015empirical}, multiple factors associated with test cases and recent modifications \cite{strandberg2016experience}, service composition interactions \cite{srikanth2016test}, or domain-specific heuristics \cite{marijan2013test}. Even though diversity-based TCP has been proposed and used previously, it has not been widely studied in combination with HBTP and in CI environments. Hemmati et al. \cite{hemmati2017prioritizing} previously investigated a history-based diversity approach in Firefox project. They used previous failure knowledge in combination with a single similarity metric, Manhattan distance, using the English texts of manual test cases. In contrast, we investigated history-based diversity using three different similarity metrics (Manhattan, NCD, and NCD Multiset) with different history interval sizes, over a large number of automated builds extracted from six open-source projects. This has not been studied in the past, to the best of our knowledge. 

Third, for the history-based diversity approach to be applicable in a CI environment, both effectiveness and performance are critical. In this study, apart from effectiveness, we investigated the method execution time within and across the studied projects.

%% ********************************************************************************************
%% CHAPTER 3. RESEARCH METHOD
%% ********************************************************************************************
\section{Research Method}
The study's objective and research questions, data properties and data extraction procedure, study design, and analysis methods are discussed, in that order.

\subsection{Objective and Research Questions}
The main objective of our study is to shorten the RT feedback cycle for continuous integration of software systems. In other words, we aim to catch regression faults earlier, allowing developers to integrate and verify their changes more frequently and continuously. To achieve that end, we investigated six open-source software projects, each of which included several builds over a large period of time. The research questions and their rationales are as follows:

\begin{itemize}
\item \textbf{RQ1: To what extent can regression faults be captured effectively by using previous failure knowledge?} This research question is designed to analyze regression faults according to their past verdicts and study the extent to which they can be captured using knowledge of their previous failures.

\item \textbf{RQ2: Does the effectiveness of HBTP change over time with a larger history interval?} This research question is designed to study whether the effectiveness of HBTP changes over time if a larger history interval size is used. In this question, we perform HBTP using a differing number of previous verdicts and investigate whether its effectiveness changes by varying the size of the history interval.

\item \textbf{RQ3: Does the effectiveness of HBTP change when combined with diversity-based prioritization?} This research question is designed to study whether the effectiveness of HBTP changes by re-ordering tests based on their distances (i.e., their dissimilarity) to the set of already prioritized tests.

\item \textbf{RQ4: Among the investigated test prioritization techniques, which is most effective and has the best performance compared to the others?} This research question is designed to compare the effectiveness and performance of investigated TCP techniques within and across the studied subjects.
\end{itemize}

\subsection{Software Subjects}

To perform our experiment, we used TravisTorrent \cite{msr17challenge}, a freely available database based on Travis CI and GitHub, that provides access to the build information of several projects. TravisTorrent\textsc{\char13}s database includes several properties and information about a project\textsc{\char13}s Travis build history \footnote{A list of data properties provided by TravisTorrent is provided here:
\url{https://travistorrent.testroots.org/page_dataformat/}.}. However, TravisTorrent\textsc{\char13}s database does not include information about the list of executed tests or their verdicts (i.e., passed or failed) for each revision. To extract the required data for our study, we downloaded all available build logs for the studied projects and automatically analyzed them \footnote{To enable a deeper analysis, TravisTorrent also provides raw build logs here: \url{https://travistorrent.testroots.org/buildlogs/}}. The following data properties were extracted:

\begin{itemize}
\item Commit-ID: The unique identifier of the original commit extracted from TravisTorrent. 
\item Files modified: The list of modified files, extracted from GitHub for a particular commit. 
\item Test results: The list of executed tests and their verdicts, extracted from build logs. 
\item Build time: The original build time, extracted from TravisTorrent. 
\end{itemize}

The build logs typically include a list of executed test files and their verdicts. Therefore, the granularity level in this study is at the file level, although the investigated prioritization techniques can be applied to any level if data is available. In practice, any software project might be a candidate subject if its build logs are available and include information on the list of executed tests and their verdicts. The structure and format of build logs varies among software projects, depending on the build and testing framework used during the development. For the purpose of this study, we selected six Java-based software projects in order to minimize the effort in the data collection phase. However, the above-mentioned data properties can be gathered easily from any CI development environment, independent of its underlying technologies. 

Table \ref{table:subj_char} represents the characteristics of analyzed projects. The first column indicates the project name and abbreviated identifier. The second column is a range and indicates the analyzed build time period. The third column shows the actual number of analyzed builds; the parentheses indicate the number of builds which included RT output. The fourth column shows the number of builds in which at least one test failed. The fifth column shows the number of unique tests identified from build logs, while the sixth column is a range that refers to the number of executed tests during builds (each of which might include several test cases). The last column shows the source line-of-code (SLOC) for the most recent studied version, as reported by SLOCCount \footnote{SLOCCount is a suite of programs used to count lines of code: \url{https://www.dwheeler.com/sloccount/}}. The abbreviated ID is used in the remaining sections when we refer to these projects.

\begin{table}[H]
\centering
\caption {Subject Characteristics}
\label{table:subj_char}
\begin{tabular}{|p{3.5cm}|p{2cm}|p{1cm}|p{1cm}|p{0.75cm}|p{1.5cm}|p{1.25cm}|}
\hline                       
  Name (ID) &  Time Frame & Builds & Faulty Builds & Tests & Test Suite Size & SLOC\\
\hline  
\hline 
Google Guava (GUV) & 2014/11/05-2016/08/29 & 465 (458) & 41 & 411 & 365-404 & 247,497\\
\hline 
MyBatis (MYB) & 2013/02/14-2016/08/23 & 988 (923) & 27 & 278 & 198-241 & 86,549\\
\hline  
Apache Tajo (TAJ) & 2014/05/08-2016/08/29 & 4670 (3908) & 564 & 313 & 30-241 & 248,155\\
\hline    
AWS Java SDK (AWS) & 2013/05/01-2016/08/31 & 863 (449) & 85 & 144 & 58-114 & 1,411,875\\
\hline   
DSpace (DSP) & 2013/07/25-2016/08/31 & 3813 (3327) & 71 & 80 & 16-61 & 289,703\\
\hline   
Apache Storm (STM) & 2015/04/23-2016/09/01 & 2196 (1962) & 542 & 135 & 6-60 & 214,437\\
\hline   
\end{tabular}
\end{table}

\subsection{Study Design}
\label{section_studydesign}

\subsubsection{RQ1: To what extent can regression faults be captured effectively by using previous failure knowledge?}

The objective of this research question was to investigate the extent to which regression faults can be captured using previous failure knowledge. To answer this question, we classified regression faults according to their past verdicts, i.e., those that can be captured using previous failure knowledge (T1) and those without any previous failure (T2). In other words, T1 includes regression faults that are detected by test cases which have failed earlier (i.e., failure status observed in the earlier execution of the test case). In contrast, T2 includes faults that are detected by test cases which have never failed earlier (i.e., no previous verdicts exist or previous execution of the test case have been always passed). This information might be helpful in understanding the nature of regression faults in the investigated projects, e.g., whether they can be predicted and captured using previous failure knowledge. 

For T1 regression faults, we further calculated the gap (number of verdicts) between the observed failure and the previous failure. This information helps us to understand the distances among regression faults and might be useful in adjusting the interval size of HBTP, that is, the number of previous verdicts we should take to capture regression faults effectively. For T2 regression faults, we calculated the gap (number of verdicts) between an observed failure and the first available verdict. This information might be an indication of the age of fault-revealing tests within our historical data.

\subsubsection{RQ2: Does the effectiveness of HBTP change over time with a larger history interval?}

The objective of this research question was to investigate whether the effectiveness of HBTP changes by varying the size of the history interval (i.e., the number of past verdicts taken into account). The effectiveness of HBTP was measured using the average percentage of faults detected (APFD), which is a common metric used in the literature and elaborated in section \ref{section_evaluation}. The HBTP used in this study is similar to the cluster-based technique proposed by Hemmati et al. \cite{hemmati2017prioritizing}. The rationale is that such an approach only requires previous failure knowledge and that we have access to such information. 

In our study, we calculated the cumulative priority for each test using its previous failures over the last $N$ builds (depending on the interval size). The highest weight corresponds to the failure exposed in a previous build (current$-1$) and the failure in every preceding build is weighted lower than the failure in its successive build. Specifically, failures are weighted by their distance $W_n$, that reflect the impact of the failure occurred in the past and is $n$ builds far from the current build session. The following values were assigned to the weights.

\begin{equation*}
    W_n = \begin{cases}
               0.9, & if \quad n = 1 \\
               0.8, & if \quad n = 2 \\
               0.7, & if \quad n = 3 \\
               0.6, & if \quad n = 4 \\
               0.5, & if \quad n = 5 \\
               0.4, & if \quad n = 6 \\
               0.3, & if \quad n = 7 \\
               0.2, & if \quad n = 8 \\
               0.1, & if \quad n \geq 9 \\
           \end{cases}
\end{equation*}

Thereafter, we aggregated tests into clusters based on their weight and sorted these clusters in descending order. To answer this research question, we compared the effectiveness of HBTP with random permutation (RND) and with each other using different interval sizes (V1, V10, V100, V500). Using HTBP approach, test cases without historical failure are grouped in a single cluster and remain in their original order (i.e, the order in which they executed and appeared in the build logs). There was also the possibility that the investigated projects might have already used TCP techniques. To avoid the impact of this and create a fair comparison with random ordering, we simply randomized the intra-cluster tests for the purpose of this research question. This technique is called history-based random (HBR) in our study.

Coverage-based TCP is an important baseline in the literature, but excluded in our experiment due to the difficulties associated with collecting coverage data for the subject programs. The coverage data can be on different level (e.g., statement, branch, method coverage) and obtained using dynamically analyzing the program execution or by statically analyzing the test and source-code. Gathering coverage data was challenging with our subjects, because test information have been extracted from archival data (build logs), and not from actual test executions. Furthermore, the high frequency of changes in CI environment quickly makes the data gathered by code instrumentation imprecise and obsolete \cite{elbaum2014techniques}. The challenges associated with coverage-based TCP in CI environments are discussed in the literature, e.g., by Elbaum et al. \cite{elbaum2014techniques} and Hemmati et al. \cite{hemmati2017prioritizing}. Future studies are required to address these challenges and  develop efficient instrumentation tools for CI environment.

\subsubsection{RQ3: Does the effectiveness of HBTP change when combined with diversity-based prioritization?}

Due to the lack of previous failure knowledge, many tests in our approach cannot be effectively prioritized. These tests end up in a single cluster if we solely prioritize them based on previous failure knowledge. This includes both newly added tests or those which have not revealed any failure in previous builds. Furthermore, within each cluster, there might be several tests with the same weight. To break the tie, the intra-cluster tests can be randomized or other TCP techniques can be employed in combination with HBTP. 

The main objective of this research question was to investigate whether the effectiveness of HBTP changes when it is combined with diversity-based prioritization. To answer this research question, we studied history-based diversity (HBD) over different interval sizes and compared its effectiveness with HBR. The latter technique simply randomizes tests within each cluster. In contrast, HBD iteratively prioritizes the intra-cluster tests on the basis of their distance (dissimilarity) to the set of already prioritized tests. Apart from the interval sizes used in RQ2 (V1, V10, V100, V500), we also studied the initial stage (V0), where there was no historical data available at hand and the test suite was completely prioritized using test distances.

To answer this research question, we implemented HBD using three similarity metrics, including Manhattan, NCD, and NCD Multiset. These approaches were selected because of their promising results reported in recent studies \cite{ledru2012prioritizing,henard2016comparing,feldt2016test}. To calculate the distances, we automatically downloaded source code from Github for all studied revisions and used the source code behind the tests at their exact revision \footnote{Github allows offline access to all revisions: \url{https://github.com/{username}/{projectname}/archive/{sha}.zip}}. We implemented the Manhattan and NCD approaches using a pairwise algorithm proposed by Ledru et al. \cite{ledru2012prioritizing}. For the NCD Multiset, we implemented the following algorithm, which iteratively picks a test that has maximum distance (is most dissimilar) to the entire set of already prioritized tests (rather than between pairs). Note that $C$ is a function that calculates the approximate Kolmogorov complexity and returns the length of the input string after its compression, using a chosen compression program. In this study, we used LZ4, which is a high-speed lossless data compression algorithm and is widely used in search engines and database management systems \footnote{The LZ4 compression algorithm and its implementation: \url{http://lz4.github.io/lz4/}}.\\

\begin{algorithm}[H]
\SetAlgoLined
\KwData{Test Suite $T$ and Prioritized Set $PS$}
\KwResult{$PS$}
 \While{$T$ is not empty}{
  Find $T_i$ which maximizes $C(PS, T_i)$\;
  Append $T_i$ to $PS$\;
  Remove $T_i$ from $T$\;
 }
 \caption{Test Prioritization Using NCD Multiset}
\end{algorithm}

\subsubsection{RQ4: Among the investigated test prioritization techniques, which is most effective and has the best performance compared to the others?}

For the HBD approach to be applicable in a CI context, both effectiveness and performance are critical. Thus, apart from assessing the effectiveness in terms of APFD, we studied the performance of HDB techniques in terms of method execution time. During the initial stages of this research, we observed that calculating distances constitutes the largest portion of method execution time. To improve the performance of HBD, we implemented a simple caching system where distances among the tests are calculated upon request and retained. The distance value is updated only if one of the relevant tests is observed in the change list. This approach greatly reduced the method execution time when we prioritized tests over several revisions using pairwise algorithms (NCD and Manhattan). However, such a caching system is ineffective with the NCD Multiset algorithm because the distance is calculated at the level of entire sets of elements rather between pairs. To speed up the process, we parallelized both pairwise and multiset algorithms using concurrent utilities proposed by Java specification request (JSR) 166. The above-mentioned methods greatly improved the performance of our HBD techniques.

\subsection{Evaluation}
\label{section_evaluation}
To assess the effectiveness of TCP techniques, we used APFD metric that was originally introduced by Rothermel et al. \cite{rothermel2001prioritizing} and is widely used in the literature. Let $T$ be an ordered test suite, containing $n$ test cases, and $F$ be a set of $m$ faults detected by $T$; then $TF_{i}$ indicates the number of test cases executed in $T$ before capturing fault $i$. APFD indicates the average percentage of faults detected and is defined as follows:

$$ APFD = 100 * (1- \frac{ TF_1 + TF_2 + ... + TF_M } { nm } + \frac{1} { 2_n }) $$

In order to properly compare the investigated TCP techniques, we conducted statistical analyses. The Mann\textendash Whitney U test \cite{arcuri2011practical}, a non-parametric significance test, was applied to determine whether the difference between two compared techniques was statistically significant (p-value is less than 0.05). The null hypothesis of this test indicates that there is no significant difference between APFDs of the two techniques under evaluation. The Mann\textendash Whitney U test was selected because the studied data may not follow normal distribution. The significance test indicates whether the difference between two compared techniques is statistically significant, but does not show the size of the difference between them. Thus, we used the Vargha\textendash Delaney A measure \cite{arcuri2011practical}, which is a non-parametric effect size. The Vargha\textendash Delaney A measure is a number between 0 and 1. When the A measure is 0.5, the two compared techniques, X and Y, are equal. When the A measure is higher than 0.5, it means that X outperformed Y, and vice versa. Furthermore, when comparing TCP techniques, we also provided violin plots to visualize the distribution of APFDs.

%% ********************************************************************************************
%% CHAPTER 4. RESULTS
%% ********************************************************************************************
\section{Findings}
This section is structured to address the research questions and includes the aggregated analysis of results from our experiments. The experiments were conducted on a computer with Intel 2.7 GHz Xeon E5-2680 (8 cores) and 16 GB installed RAM.

\subsection{RQ1: To what extent can regression faults be captured effectively by using previous failure knowledge?}

Table \ref{table:fault_char} shows the aggregated number of fault-revealing tests, regression faults, and the ratio of two different types of regression faults. It can be seen that in the majority of investigated projects, only a small proportion of tests has ever failed (less than 11\% in four projects, and 3\textendash 52\% overall). This indeed underscores the importance of TCP in CI environments, where RT is performed more frequently and continuously. From a historical perspective, regression faults can be classified into two groups: those that can be captured using previous failure knowledge (T1), and those without any previous failure (T2). Our findings indicate that the majority of regression faults among all investigated projects are T1 (57\textendash 97\%) and can be captured solely by using previous failure knowledge. In four of the investigated subjects, only a small proportion (less than 13\%) of regression faults are classified as T2 and cannot be captured using previous failure knowledge. This number is higher in the MYB (43\%) and GUV (23\%) projects.

\begin{table}[H]
\centering
\caption {Fault-revealing Tests and Regression Faults}
\label{table:fault_char}
\begin{tabular}{|l|l|l|l|l|l|}
\hline                       
 Project & \# Tests & \# Fault-revealing Tests & \# Faults & T1 Faults & T2 Faults \\
\hline
\hline 
MYB & 278 & 26 (9.3\%) & 60 & 57\% & 43\% \\
\hline  
GUV & 411 & 13 (3.1\%) & 56 & 77\% & 23\% \\
\hline
AWS & 144 & 14 (9.7\%) & 123 & 89\% & 11\% \\
\hline
DSP & 80 & 42 (52.5\%) & 353 & 88\% & 12\% \\
\hline
STM & 135 & 14 (10.3\%) & 564 & 97\% & 3\% \\
\hline
TAJ & 313 & 136 (43.4\%) & 1087 & 87\% & 13\% \\
\hline
\end{tabular}
\end{table}

To gain a better understanding of the regression faults, we conducted a deeper analysis. For T1 regression faults, we calculated the gap (number of verdicts) between the observed failure and the previous failure. For T2 regression faults, we calculated the gap between the observed failure with the first available verdict. Table \ref{table:five_num} shows the five-number summary of gaps for T1 and T2 regression faults. The gap between two observed failures varies among the studied projects and can be used to adjust the effectiveness of HBTP. From the results, it can be seen that 50\% of T1 regression faults within three investigated projects (MYB, GUV, and AWS) can be captured only by using the last verdict (current$-1$). For the AWS project, only 16 previous verdicts are required to capture 100\% of T1 regression faults, which constitute 89\% of total regression faults. For T2 regression faults, a small portion of regression faults occurred when a test was introduced to the test suite (25\% in the case of AWS). However, the majority of T2 faults were introduced due to the change in the system and previous failure knowledge was not available. Thus, extra information is required to effectively capture these regression faults.

\begin{table} [H]
\centering
\caption{Regression Faults and Gaps \textendash{} 5-number summary}
\label{table:five_num}
  \begin{tabular}{|l|l|l|l|l|l|l|l|l|l|l|}
    \hline
    \multirow{2}{*}{Project} &
      \multicolumn{5}{c|}{T1 Regression Faults} &
      \multicolumn{5}{c|}{T2 Regression Faults} \\
    & Min & P25 & P50 & P75 & Max & Min & P25 & P50 & P75 & Max \\
    \hline
    \hline
    MYB & 1 & 1 & 1 & 94 & 268 & 0 & 114 & 192 & 289 & 900\\
    \hline
    GUV & 1 & 1 & 1 & 1 & 162 & 0 & 53 & 127 & 371 & 376 \\
	\hline
	AWS & 1 & 1 & 1 & 1.5 & 16 & 0 & 0 & 79 & 367 & 384 \\
	\hline
	DSP & 1 & 2 & 26 & 251 & 1023 & 18 & 440 & 748 & 748 & 1989 \\
	\hline
	STM & 1 & 1 & 3 & 6 & 447 & 0 & 4.75 & 17.5 & 191 & 600 \\
	\hline
	TAJ & 1 & 3 & 18 & 108 & 2243 & 0 & 147 & 650 & 1559 & 3241 \\
	\hline	
  \end{tabular}
\end{table}

\subsection{RQ2: Does the effectiveness of HBTP change over time with a larger history interval?}

For each faulty build within the investigated projects, we prioritized tests using different history interval sizes, i.e., the number of previous verdicts (V1, V10, V100, V500). The findings presented in this section are based on the aggregated results of all execution rounds. For each project, we had access to a different number of faulty builds. Thus, the aggregated results are based on different numbers of observations (see the number of faulty builds presented in Table \ref{table:subj_char}).

\begin{figure}[h]
\centering
\includegraphics[width=1\columnwidth]{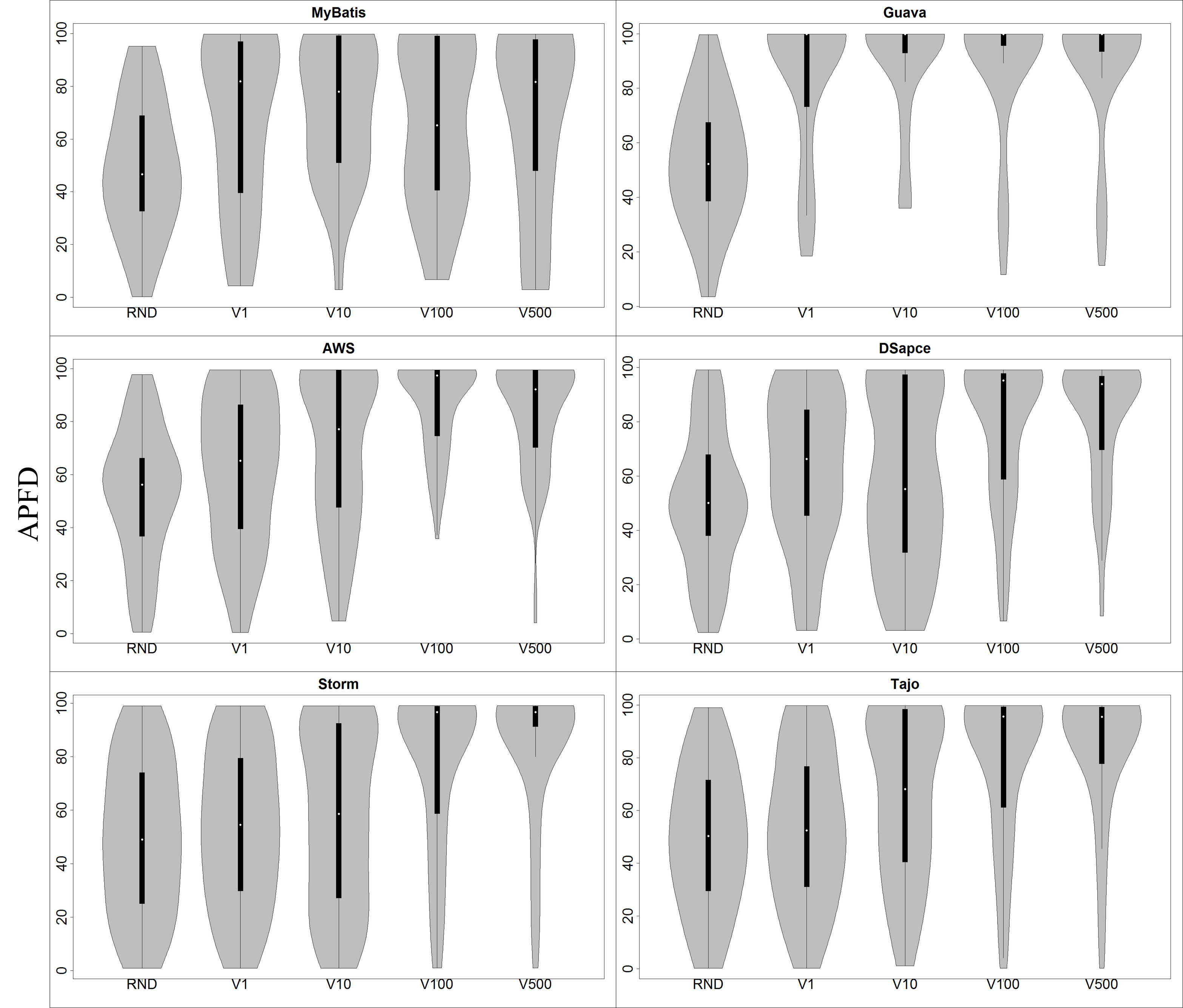}
\caption{RND vs. HBR over different interval sizes (V)} \label{fig:rq2}
\end{figure}

Table \ref{table:rq2} shows the effectiveness of HBTP using different interval sizes. It can be seen that HBTP does not necessarily need to have a large amount of historical data. Even with the last verdict (current$-1$), improvement (0.53\textendash 0.82) was observed in all projects in comparison to random ordering. When we took the last 10 verdicts, greater improvement (0.53\textendash 0.61) was  observed in five studied projects in comparison to HBTP-V1. When we extended our interval size to the last 100 verdicts, the effectiveness of HBTP improved (0.61\textendash 0.67) in four investigated projects. Taking the last 500 verdicts would lead to negligible improvement (0.51 and 0.54) in only two subjects. The remaining projects show a negligible decline (0.47\textendash 0.49) on APFD. Our findings imply that the effectiveness of HBTP changes over time by taking a larger history interval. However, the impact of the phenomenon varies among projects and is perhaps associated with the nature of regression faults and their distances (see RQ1). Overall, within the investigated projects, we observed that, to a certain degree, changing the interval size would lead to positive improvement in terms of APFD. Figure \ref{fig:rq2} illustrates the full distribution of APFDs for RND and HBR over different interval sizes.

\begin{table} [H]
\centering
\caption {Effectiveness comparison - RND vs. HBR over different interval sizes (V)}
\label{table:rq2}
  \begin{tabular}{|l|l|l|l|l|l|l|l|l|l|}
    \hline
    \multirow{2}{*}{Project} &
      \multicolumn{5}{c|}{Mean} &
      \multicolumn{4}{c|}{Effect-size (X vs- Y)} \\
    & RND & V1 & V10 & V100 & V500 & V1-RND & V10-1 & V100-10 & V500-100 \\
    \hline
    \hline
	MYB&48.93&67.14&71.13&66.44&67.39&0.68\cellcolor{Gray}&0.53&0.44&0.48 \\
	\hline
	GUV&53.15&84.45&88.95&87.85&88.32&0.82\cellcolor{Gray}&0.54&0.46&0.49 \\
	\hline
	AWS&52.57&62.01&69.64&86.01&83.32&0.60\cellcolor{Gray}&0.58&0.64\cellcolor{Gray}&0.47 \\
	\hline
	DSP&50.86&62.80&58.64&77.99&81.48&0.62\cellcolor{Gray}&0.47&0.64\cellcolor{Gray}&0.48 \\
	\hline
	STM&49.75&53.24&57.81&77.68&84.19&0.53\cellcolor{Gray}&0.54\cellcolor{Gray}&0.67\cellcolor{Gray}&0.54\cellcolor{Gray} \\
	\hline
	TAJ&49.51&53.59&65.62&78.85&83.38&0.53\cellcolor{Gray}&0.61\cellcolor{Gray}&0.61\cellcolor{Gray}&0.51 \cellcolor{Gray}\\
	\hline
  \end{tabular}
\end{table}

\subsection{RQ3: Does the effectiveness of HBTP change when combined with diversity-based prioritization?}

To answer this question, we compared the effectiveness of HBR with HBD over different interval sizes. The latter technique prioritizes the intra-cluster tests based on their distance (dissimilarity) to the set of already prioritized tests. In contrast, HBR simply randomizes tests within each cluster. Three different similarity metrics were used, including NCD, NCD Multiset (NCD-MS), and Manhattan (MNH). Apart from the interval sizes used in RQ2, we also studied the initial stage (V0), when there is no historical data available at hand, and the test suite is completely prioritized using test distances. Table \ref{table:rq3_p1} to \ref{table:rq3_p6} present the APFD mean and effect size of different techniques for each project using different interval sizes. Figures \ref{fig:RQ3_1} to \ref{fig:RQ3_6} presented in \ref{appendix_plots} show the violin plots for each project over different interval sizes. 

The results indicate a noticeable improvement when no historical data was available (V0) and the test suite was prioritized solely by using test distances. This implies that diversity-based TCP can be used effectively during the early stages of HBTP deployment, when no historical data is available. Among the investigated techniques, NCD-MS was the best and greatly improved the effectiveness of HBTP-V0 (0.68\textendash 0.91) in comparison to MNH (0.47\textendash{} 0.91) and NCD (0.51\textendash 0.92). When we combined the diversity-based approach with HBTP (V1, V10, V100, V500), positive improvement was also observed in the majority of cases (19/24 cases for NCD and Manhattan, and 24/24 for NCD-MS). Among the investigated techniques, NCD-MS was the best and consistently improved the effectiveness of HBTP (0.51\textendash 0.73) using different interval sizes. These differences are also statistically significant in the majority of the cases for NCD-MS.

Our findings indicate that the effectiveness of HBTP changes when combined with diversity-based TCP. In other words, re-ordering intra-cluster tests based on their dissimilarity to the set of already prioritized tests has a positive impact on the effectiveness of HBTP. This impact seems to be relatively smaller when we take a larger history interval into account. This is perhaps due to the fact that taking more verdicts leads to more clusters, with a smaller number of intra-cluster tests. Thus, more regression faults can be potentially captured and there is a higher chance of a draw between HBTP and HBD. On the other hand, more clusters of a smaller size means that less flexibility is given to the diversity-based approach in re-ordering these intra-cluster tests. The violin plots presented in \ref{appendix_plots} show that HBD not only improved effectiveness, but also tended to have less variance in the results.

\begin{table} [H]
\centering
\caption {MYB\textendash HBD vs. HBR over different interval sizes (V)}
\label{table:rq3_p1}
  \begin{tabular}{|l|l|l|l|l|l|l|l|}
    \hline
    \multirow{2}{*}{MYB} &
      \multicolumn{4}{c|}{APFD Mean} &
      \multicolumn{3}{c|}{Effect-size X vs. HBR} \\
    & HBR & NCD & NCD-MS & MNH & NCD & NCD-MS & MNH \\
    \hline
    \hline
    V0&45.13&66.94&69.62&66.37&0.72\cellcolor{Gray}&0.74\cellcolor{Gray}&0.72\cellcolor{Gray}\\
	\hline
	V1&67.14&77.6&80.32&77.55&0.57&0.61&0.57\\
	\hline
	V10&71.13&78.92&81.04&77.95&0.57&0.59&0.56\\
	\hline
	V100&66.44&78.41&80.6&75.25&0.58&0.61\cellcolor{Gray}&0.57\\
	\hline
	V500&67.39&78.71&81.12&76.33&0.55&0.58\cellcolor{Gray}&0.55\\
	\hline
  \end{tabular}
\end{table}

\begin{table} [H]
\centering
\caption {GUV\textendash HBD vs. HBR over different interval sizes (V)}
\label{table:rq3_p2}
  \begin{tabular}{|l|l|l|l|l|l|l|l|}
    \hline
    \multirow{2}{*}{GUV} &
      \multicolumn{4}{c|}{APFD Mean} &
      \multicolumn{3}{c|}{Effect-size X vs. HBR} \\
    & HBR & NCD & NCD-MS & MNH & NCD & NCD-MS & MNH \\
    \hline
    \hline
	V0&48.01&87.65&84.58&83.03&0.92\cellcolor{Gray}&0.91\cellcolor{Gray}&0.91\cellcolor{Gray}\\
	\hline
	V1&84.45&93.84&94.58&94.5&0.55\cellcolor{Gray}&0.53\cellcolor{Gray}&0.54\cellcolor{Gray}\\
	\hline
	V10&88.95&94.06&95.00&94.92&0.51\cellcolor{Gray}&0.51\cellcolor{Gray}&0.52\cellcolor{Gray}\\
	\hline
	V100&87.85&93.8&94.89&94.75&0.51\cellcolor{Gray}&0.51\cellcolor{Gray}&0.51\cellcolor{Gray}\\
	\hline
	V500&88.32&93.51&94.88&94.78&0.52\cellcolor{Gray}&0.52\cellcolor{Gray}&0.52\cellcolor{Gray}\\
	\hline
  \end{tabular}
\end{table}

\begin{table} [H]
\centering
\caption {AWS\textendash HBD vs. HBR over different interval sizes (V)}
\label{table:rq3_p3}
  \begin{tabular}{|l|l|l|l|l|l|l|l|}
    \hline
    \multirow{2}{*}{AWS} &
      \multicolumn{4}{c|}{APFD Mean} &
      \multicolumn{3}{c|}{Effect-size X vs. HBR} \\
    & HBR & NCD & NCD-MS & MNH & NCD & NCD-MS & MNH \\
    \hline
    \hline
	V0&49.65&51.35&68.56&49.29&0.51&0.70\cellcolor{Gray}&0.47\\
	\hline
	V1&62.01&58.87&72.5&55.1&0.45&0.60\cellcolor{Gray}&0.42\cellcolor{Gray}\\
	\hline
	V10&69.64&69.08&77.94&65.16&0.48&0.56\cellcolor{Gray}&0.45\\
	\hline
	V100&86.01&87.09&87.92&86.22&0.52&0.53&0.50\\
	\hline
	V500&83.32&86.92&87.69&86.22&0.54\cellcolor{Gray}&0.55\cellcolor{Gray}&0.52\\
	\hline

  \end{tabular}
\end{table}

\begin{table} [H]
\centering
\caption {DSP\textendash HBD vs. HBR over different interval sizes (V)}
\label{table:rq3_p4}
  \begin{tabular}{|l|l|l|l|l|l|l|l|}
    \hline
    \multirow{2}{*}{DSP} &
      \multicolumn{4}{c|}{APFD Mean} &
      \multicolumn{3}{c|}{Effect-size X vs. HBR} \\
    & HBR & NCD & NCD-MS & MNH & NCD & NCD-MS & MNH \\
    \hline
    \hline
	V0&52.05&54.94&69.63&68.55&0.51&0.68\cellcolor{Gray}&0.65\cellcolor{Gray}\\
	\hline
	V1&62.8&61.25&72.16&71.56&0.47&0.58\cellcolor{Gray}&0.58\cellcolor{Gray}\\
	\hline
	V10&58.64&67.45&75.65&75.36&0.57\cellcolor{Gray}&0.64\cellcolor{Gray}&0.63\cellcolor{Gray}\\
	\hline
	V100&77.99&78.04&83.09&83.03&0.49&0.53&0.53\\
	\hline
	V500&81.48&82.6&85.39&84.98&0.51&0.53\cellcolor{Gray}&0.52\cellcolor{Gray}\\
	\hline
  \end{tabular}
\end{table}

\begin{table} [H]
\centering
\caption {STM\textendash HBD vs. HBR over different interval sizes (V)}
\label{table:rq3_p5}
  \begin{tabular}{|l|l|l|l|l|l|l|l|}
    \hline
    \multirow{2}{*}{STM} &
      \multicolumn{4}{c|}{APFD Mean} &
      \multicolumn{3}{c|}{Effect-size X vs. HBR} \\
    & HBR & NCD & NCD-MS & MNH & NCD & NCD-MS & MNH \\
    \hline
    \hline
	V0&51.23&72.95&77.47&68.54&0.73\cellcolor{Gray}&0.76\cellcolor{Gray}&0.67\cellcolor{Gray}\\
	\hline
	V1&53.24&73.48&77.84&69.08&0.7\cellcolor{Gray}&0.73\cellcolor{Gray}&0.65\cellcolor{Gray}\\
	\hline
	V10&57.81&75.2&79.55&71.52&0.64\cellcolor{Gray}&0.67\cellcolor{Gray}&0.61\cellcolor{Gray}\\
	\hline
	V100&77.68&83.11&84.03&78.38&0.52\cellcolor{Gray}&0.52\cellcolor{Gray}&0.50\\
	\hline
	V500&84.19&88.15&90.68&86.72&0.5\cellcolor{Gray}&0.51\cellcolor{Gray}&0.50\cellcolor{Gray}\\
	\hline
  \end{tabular}
\end{table}

\begin{table} [H]
\centering
\caption {TAJ\textendash HBD vs. HBR over different interval sizes (V)}
\label{table:rq3_p6}
  \begin{tabular}{|l|l|l|l|l|l|l|l|}
    \hline
    \multirow{2}{*}{TAJ} &
      \multicolumn{4}{c|}{APFD Mean} &
      \multicolumn{3}{c|}{Effect-size X vs. HBR} \\
    & HBR & NCD & NCD-MS & MNH & NCD & NCD-MS & MNH \\
    \hline
    \hline
	V0&50.12&65.43&69.12&65.86&0.66\cellcolor{Gray}&0.70\cellcolor{Gray}&0.66\cellcolor{Gray}\\
	\hline
	V1&53.59&68.06&71.33&68.38&0.64\cellcolor{Gray}&0.67\cellcolor{Gray}&0.64\cellcolor{Gray}\\
	\hline
	V10&65.62&74.09&77.43&74.25&0.56\cellcolor{Gray}&0.59\cellcolor{Gray}&0.56\cellcolor{Gray}\\
	\hline
	V100&78.85&82.69&85.34&84.28&0.52\cellcolor{Gray}&0.53\cellcolor{Gray}&0.53\cellcolor{Gray}\\
	\hline
	V500&83.38&85.58&87.76&87.48&0.51\cellcolor{Gray}&0.52\cellcolor{Gray}&0.52\cellcolor{Gray}\\
	\hline
  \end{tabular}
\end{table}

\subsection{RQ4: Among the investigated test prioritization techniques, which is most effective and has the best performance compared to the others?}

For the HBD approach to be applicable in a CI context, both effectiveness (in terms of APFD) and performance (in terms of average method execution time (AMET)) are critical. Table \ref{table:rq3_apfd} compares the effectiveness of investigated techniques within and across the studied projects. For each project, we calculated the sum of APFDs achieved across all interval sizes (V0\textendash V500). Table \ref{table:rq3_time}, on the other hand, compares the performance of investigated techniques within and across the studied projects.

In terms of effectiveness, the HBD techniques achieved higher scores in comparison to HBR (68.47 average APFD across all projects). Manhattan and NCD achieved very close overall scores in terms of average APFD across all projects (77.17 and 76.99, respectively). NCD Multiset was a superior technique within and across the investigated projects and achieved the highest score (81.25 average APFD across all projects). In terms of performance, HBR was the fastest and always scored a very low AMET (0.1 second). Among HDB techniques, NCD was the best in terms of AMET (4.55 seconds), followed by Manhattan and NCD Multiset (17.58 and 57.58 seconds, respectively). Overall, our findings indicate that NCD Multiset outperforms both pairwise NCD and Manhattan in terms of effectiveness but has relatively higher overhead in terms of average method execution time (approximately 3.2 times higher than Manhattan and 12.6 times higher than NCD).

\begin{table}[H]
\centering
\caption {Effectiveness Comparison\textemdash APFD}
\label{table:rq3_apfd}
\begin{tabular}{|l|l|l|l|l|}
\hline                       
 Project & HBR & HBD NCD & HBD NCD-MS & HBD MNH \\
\hline  
\hline
	MYB& 317.23&380.58&392.7&373.45\\
	\hline
	GUV& 397.58&462.86&463.93&461.98\\
	\hline
	AWS& 350.63&353.31&394.61&341.99\\
	\hline
	DSP& 332.96&344.28&385.92&383.48\\
	\hline
	STM& 324.15&392.89&409.57&374.24\\
	\hline
	TAJ& 331.56&375.85&390.98&380.25\\
	\hline
	Total APFD& 2054.11&2309.77&2437.71&2315.39\\
	\hline
	Average APFD& 68.47&76.99&81.25&77.17\\
	\hline
\end{tabular}
\end{table}

\begin{table}[H]
\centering
\caption {Performance Comparison\textemdash AMET in seconds}
\label{table:rq3_time}
\begin{tabular}{|l|l|l|l|l|}
\hline                       
 Project & HBR & HBD NCD & HBD NCD-MS & HBD MNH \\
\hline  
\hline
	MYB&0.01&4.83&17.71&13.09\\
	\hline
	GUV&0.01&16.93&288.49&60.89\\
	\hline
	AWS&0.01&1.27&1.64&1.41\\
	\hline
	DSP&0.01&0.12&1.03&0.83\\
	\hline
	STM&0.01&0.03&0.49&0.28\\
	\hline
	TAJ&0.01&4.13&36.14&28.99\\
	\hline
	Total AMET&0.06&27.31&345.5&105.49\\
	\hline
	Average AMET&0.01&4.55&57.58&17.58\\
\hline
\end{tabular}
\end{table}

%% ********************************************************************************************
%% CHAPTER 5. DISCUSSION
%% ********************************************************************************************
\section{Discussion}
We conducted this research with the objective of improving the RT feedback cycle for continuous integration of software systems. In other words, our aim was to catch regression faults earlier, allowing developers to integrate and verify their changes more frequently and continuously. To achieve that aim, we investigated six open-source software projects, each of which included several builds over a large period of time.

\subsection{Overview of Findings and Their Implications}

\textbf{RQ1: To what extent can regression faults be captured effectively by using previous failure knowledge?} To address RQ1, we aggregated the number of fault-revealing tests, regression faults, and the ratio of two different types of regression faults. Our results show that only a small proportion of tests has ever failed within the studied subjects (\textless 11\% in four projects, and 3\textendash 52\% overall). This indeed raises the importance of TCP in CI environments, where RT is performed more frequently and continuously. We classified regression faults according to their past verdicts, i.e., those that can be captured using previous failure knowledge (T1), and those without any previous failure (T2). Our findings indicate that the majority of regression faults (57\textendash 97\%) among all investigated projects can be captured solely by using previous failure knowledge. This implies that previous failure knowledge seems to have strong predictive power in CI environments and can be used to prioritize tests effectively. 

TCP using previous failure knowledge in CI environments has been previously proposed \cite{hemmati2017prioritizing,elbaum2014techniques,marijan2013test,srikanth2016test,kim2002history,strandberg2016experience}; its effectiveness is perhaps mainly linked to the nature of development environment, in which developers perform automated builds at frequent, short intervals. Therefore, the tests that failed in previous builds have a much higher probability of failing again in the current build. This is in line with the experiment conducted by Hemmati et al. \cite{hemmati2017prioritizing}, where the authors argue that HBTP is particularly effective in rapid-release environments due to the recency of historical knowledge rather than other changes in the process. The gap (number of verdicts) between two regression faults varies among the projects and can be used to adjust the effectiveness of HBTP. The majority of T2 regression faults originated from old tests and were introduced due to the change in the system without having any previous failure. Thus, extra information is required to effectively capture these regression faults.

\textbf{RQ2: Does the effectiveness of HBTP change over time with a larger history interval?} To address RQ2, we investigated the effectiveness of HBTP with random ordering and with each other using different interval sizes. The HBTP method used in our study was similar to the cluster-based technique proposed by Hemmati et al. \cite{hemmati2017prioritizing} and was solely based on previous failure knowledge. Our findings show that HBTP does not necessarily need to have a large amount of historical data. Even with the last verdict (current$-1$), improvement (Vargha\textendash Delaney A measure: 0.53\textendash 0.82) was observed in all studied projects in comparison to random ordering. We also observed that the effectiveness of HBTP changes by varying the history interval size. Within the investigated subjects, our results indicate that, to a certain degree, changing the interval size leads to positive improvement in terms of APFD. However, the impact of the phenomenon varies among projects. This variation is perhaps linked with the nature of regression faults and the gaps among them. The higher the gap between two failures, the larger the interval size that is required to effectively capture regression faults. Overall, our results imply that HBTP can be deployed easily in practice, with negligible investment. The only required information is previous failure data, which can be collected easily in any CI environment, independent of the development technology used.

\textbf{RQ3: Does the effectiveness of HBTP change when combined with diversity-based prioritization?} To address RQ3, we investigated the effectiveness of HBD vs. HBR over different interval sizes. For the purpose of this research question, we implemented history-based diversity using three different similarity metrics, including Manhattan, NCD, and NCD Multiset. The Manhattan and NCD approaches were implemented using a pairwise algorithm proposed by Ledru et al. \cite{ledru2012prioritizing}, and NCD Multiset was implemented by using the algorithm described in section \ref{section_studydesign}. The effectiveness of HBTP changes when combined with diversity-based prioritization. In other words, re-ordering intra-cluster tests based on their dissimilarity to the set of already prioritized tests has a positive impact on the effectiveness of HBTP. We observe that taking larger history interval relatively reduces the magnitude of such impact. This is due to the fact that taking larger history interval leads to more clusters with a smaller number of intra-cluster tests. Thus, HBD has a less flexibility in re-ordering these intra-cluster tests and there is a higher chance of draw between HBTP and HBD. Overall, our findings imply that diversity-based prioritization can be used effectively during the early stages of software development, when historical data are not yet available or are scarce, and also combined with HBTP to improve its effectiveness.

\textbf{RQ4: Among the investigated test prioritization techniques, which is most effective and has the best performance compared to the others?} To achieve diversity-based TCP, the distance (dissimilarity) among test cases must be calculated using a particular method; this information must then be leveraged to perform TCP. To assess the dissimilarities among the tests, we used their source code; this included test input, test procedure, and assert statements. However, any source of information about test cases can be used in practice. During the initial stages of this research, we observed that calculating distances constituted the largest portion of method execution time, and hindered HBD application in the CI environment. For the HBD approach to be applicable in a CI context, both effectiveness and performance are critical. Thus, apart from assessing the effectiveness in terms of APFD, we studied the performance of HDB techniques in terms of method execution time.

Both pairwise and multiset algorithms have $O(n^{2})$ complexity and their performance is directly proportional to the square of the size of the input data. Using a pairwise algorithm, the distances among the tests can be calculated upon request, retained, and updated only if one of the relevant tests appears in the change list. However, due to the nature of the NCD Multiset algorithm, a caching system such as this is ineffective. In this study, we parallelized both algorithms using concurrent utilities proposed by JSR 166. Among the investigated HBD techniques, we found that pairwise Manhattan and NCD achieved very close overall scores in terms of average APFD across all projects (77.17 and 76.99, respectively). HBD using NCD Multiset is superior in terms of effectiveness (81.25) but comes with relatively higher overhead in terms of method execution time.

Future studies are required to investigate possible approaches to improving the performance of distance calculation for the large number of test cases. Similarity metrics like NCD have a wide range of application areas and are frequently used outside software engineering literature (see the many references in Google Scholar to \cite{li2004similarity,cilibrasi2005clustering}). To efficiently apply them to software engineering problems, the existing body of knowledge must be explored systematically and potential approaches need to be adapted. The effectiveness and performance of both the NCD pairwise and the NCD Multiset algorithms are partly linked to the chosen compression library. In this study, we used LZ4, a high-speed lossless data compression algorithm. Future research should also investigate different compression algorithms and benchmark their effectiveness and performance for the purpose of TCP.

\subsection{Threats to Validity}

In the context of empirical software engineering, validity threats are classified into four distinct categories, including construct validity, internal validity, external validity, and reliability \citep{wohlin2000experimentation}.

Threats to construct validity have to do with whether we are measuring what we intend to measure and correspond to the use of proper measures. To assess the effectiveness of TCP techniques, several metrics have been proposed in the literature (see the latest systematic literature review on TCP by Khatibsyarbini et al. \cite{khatibsyarbini2017test}). These metrics were developed to address various TCP objectives. To assess the effectiveness of TCP techniques, we used APFD; originally introduced by Rothermel et al. \cite{rothermel2001prioritizing}, this is the most frequently used metric in TCP literature \cite{khatibsyarbini2017test}. Apart from effectiveness, we also measured average method execution time in order to assess the performance of investigated techniques.

Threats to internal validity have to do with the relationship between constructs and proposed explanations. They correspond to the potential faults in our implementation, e.g., data collection, distance calculation, prioritization algorithm, and the measures used, such as APFD. To minimize the likelihood of errors in our implementation, we took several countermeasures into consideration. During the implementation phase, we followed an iterative, incremental approach, using small examples. For instance, to validate the data gathered from build logs, we randomly selected several build logs from each project and manually verified the extracted data. Furthermore, our implementation and results were discussed and reviewed in regular meetings, which were held among the co-authors of this study. In our implementation, we also strove to reuse reliable components as much as possible, i.e., libraries to calculate similarity metrics. The HBTP algorithm and the pairwise algorithm were adopted and implemented based on the existing studies \cite{hemmati2017prioritizing,ledru2012prioritizing}. The only algorithm that we designed from scratch was NCD Multiset for TCP, which has been explained in detail in this paper and is simple to implement. Overall, our implementation went through several iterations and validations before the actual experiment was conducted. The statistical analysis was done in R, using reliable packages.

Threats to external validity have to do with the generalizability of the results and correspond to whether the subjects of our study are representative of real programs. In our study, we investigated six different open-source software projects, each of which included several builds over a large period of time (see Table \ref{table:subj_char}). Thus, the subjects of our experiment are real-world projects and include real-world CI builds, test suites, and regression faults. Test suite size for a couple of the investigated projects seemed rather small, but were more comparable to enterprise-sized applications in the cases of the GUV, MYB, and TAJ projects. Test suite size was extracted from build logs and indicated the number of test files executed during the builds (each of which might include several test cases). We should also emphasize that we had access only to the specific build time period provided by TravisTorrent, and our data does not have full coverage of all revisions. Overall, our findings are valid within the investigated projects; it is difficult to generalize our results beyond the scope of the study. This motivates our future work to apply the investigated techniques in industry and to larger systems.

Apart from the APFD metric, we also measured average method execution time to assess the performance of investigated techniques. The method execution time in our study did not include test suite execution time. To perform a comprehensive assessment of the efficiency of TCP techniques, end-to-end time, i.e., TCP execution time, must be considered in addition to test suite execution time. Test suite execution time can be extracted automatically by analyzing build logs. However, such information has no relevance, as the test suite is executed in a different environment. To have a fair assessment of TCP techniques, both TCP method and test suite execution time need to be measured under the same environment. This is only possible by downloading all studied revisions from Github, building the source code, and running the test suite. Such a procedure involves manual effort and automating the entire procedure was not feasible within the scope and constraints of this study.

Threats to reliability have to do with the repeatability of the research and correspond to the possibility of reaching the same conclusion reached by the original study. Repeatability required access to the data that was used and a thorough report of the research process that was applied. The data used in our study was extracted from TravisTorrent and Github, and are publicly available. Our experiment was designed based largely on existing research \cite{ledru2012prioritizing,hemmati2017prioritizing,feldt2016test}. Nevertheless, the study design, along with the careful explanation of our implementation, were reported in this paper.

%% ********************************************************************************************
%% CHAPTER 6. CONCLUSION
%% ********************************************************************************************
\section{Concluding Remarks}

Agile software development has become a source of competitive advantage in the industry. This development paradigm calls for more frequent and continuous integration of changes; as a consequence, the demand for optimized regression testing has increased. The main objective of this research was to shorten the RT feedback cycle for continuous integration of software systems. In other words, our aim was to catch regression faults earlier, allowing developers to integrate and verify their changes more frequently and continuously. To achieve that end, we investigated six open-source software projects, each of which included several builds over a large period of time. 

In summary, the results from our experiments suggest the following: (1) Historical failure knowledge seems to have strong predictive power in CI environments and can be used to effectively prioritize test cases for execution; (2) HBTP does not necessarily require a large amount of historical data, and its effectiveness improves to a certain degree with a larger history interval; (3) Diversity-based TCP can be used effectively during the early stages of software development, when historical data are not yet available or are scarce, and also combined with HBTP to improve its effectiveness; (4) Among the investigated TCP techniques, we found that history-based diversity using NCD Multiset is superior in terms of effectiveness but comes with relatively higher overhead in terms of method execution time.

Taken together, these observations imply that HBTP can be employed in practice with negligible investment and its effectiveness can be further improved by considering distances (dissimilarities) among the tests. Our study contributes to the literature by providing empirical evidence in support of two previously proposed heuristics namely history-based and diversity-based TCP in CI environments. In the future, we plan to replicate our study in industry and to larger systems.

\section*{References}
\bibliography{mybibfile}

%% ********************************************************************************************
%% APPENDIX
%% ********************************************************************************************
\appendix 

\section{Effectiveness comparison - violin plots}
\label{appendix_plots}

\begin{figure}[H]
\centering
\includegraphics[width=1\columnwidth]{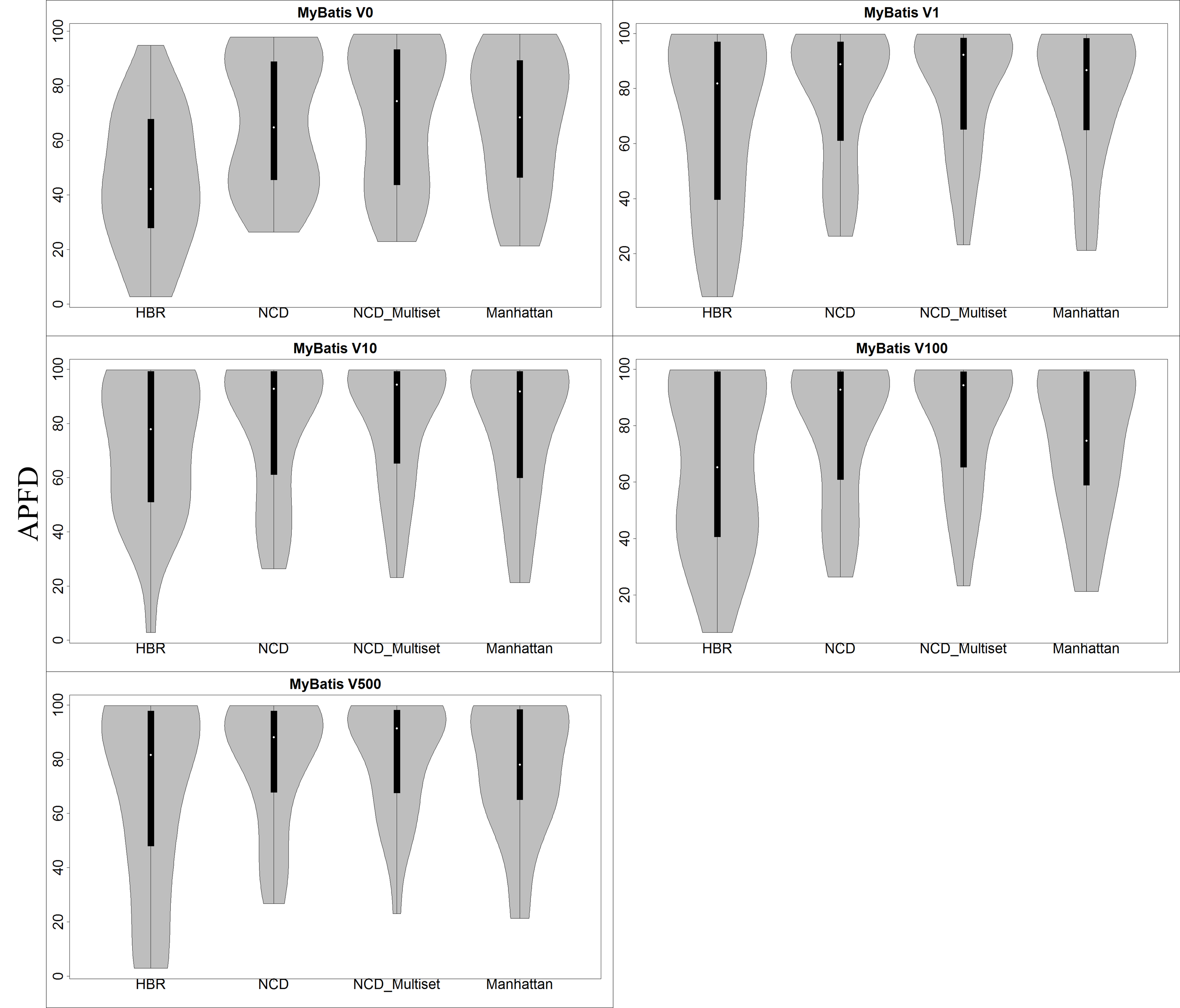}
\caption{MyBatis\textendash HBD vs. HBR over different interval sizes (V)} \label{fig:RQ3_1}
\end{figure}

\begin{figure}[H]
\centering
\includegraphics[width=1\columnwidth]{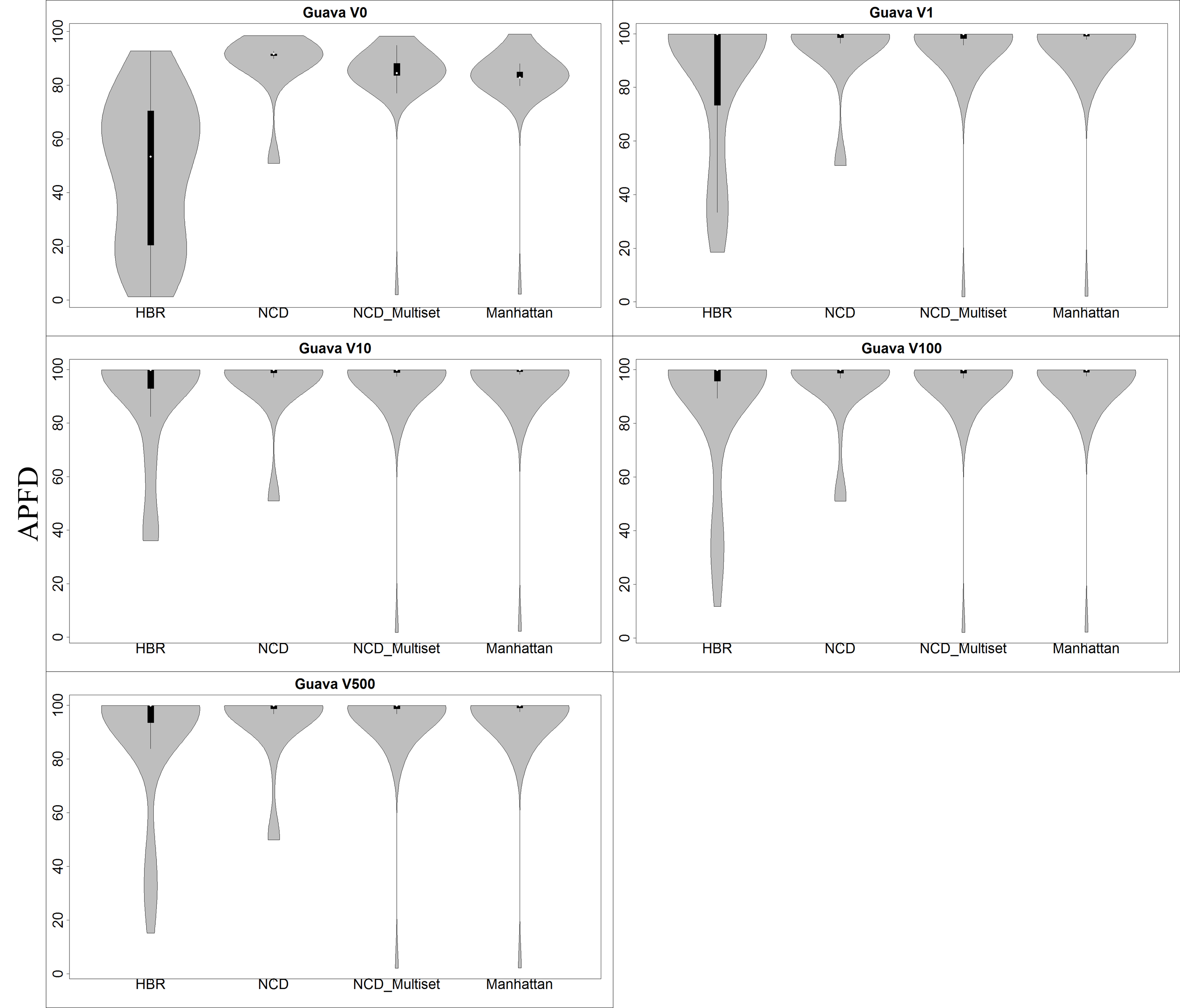}
\caption{Google Guava\textendash HBD vs. HBR over different interval sizes (V)} \label{fig:RQ3_2}
\end{figure}

\begin{figure}[H]
\centering
\includegraphics[width=1\columnwidth]{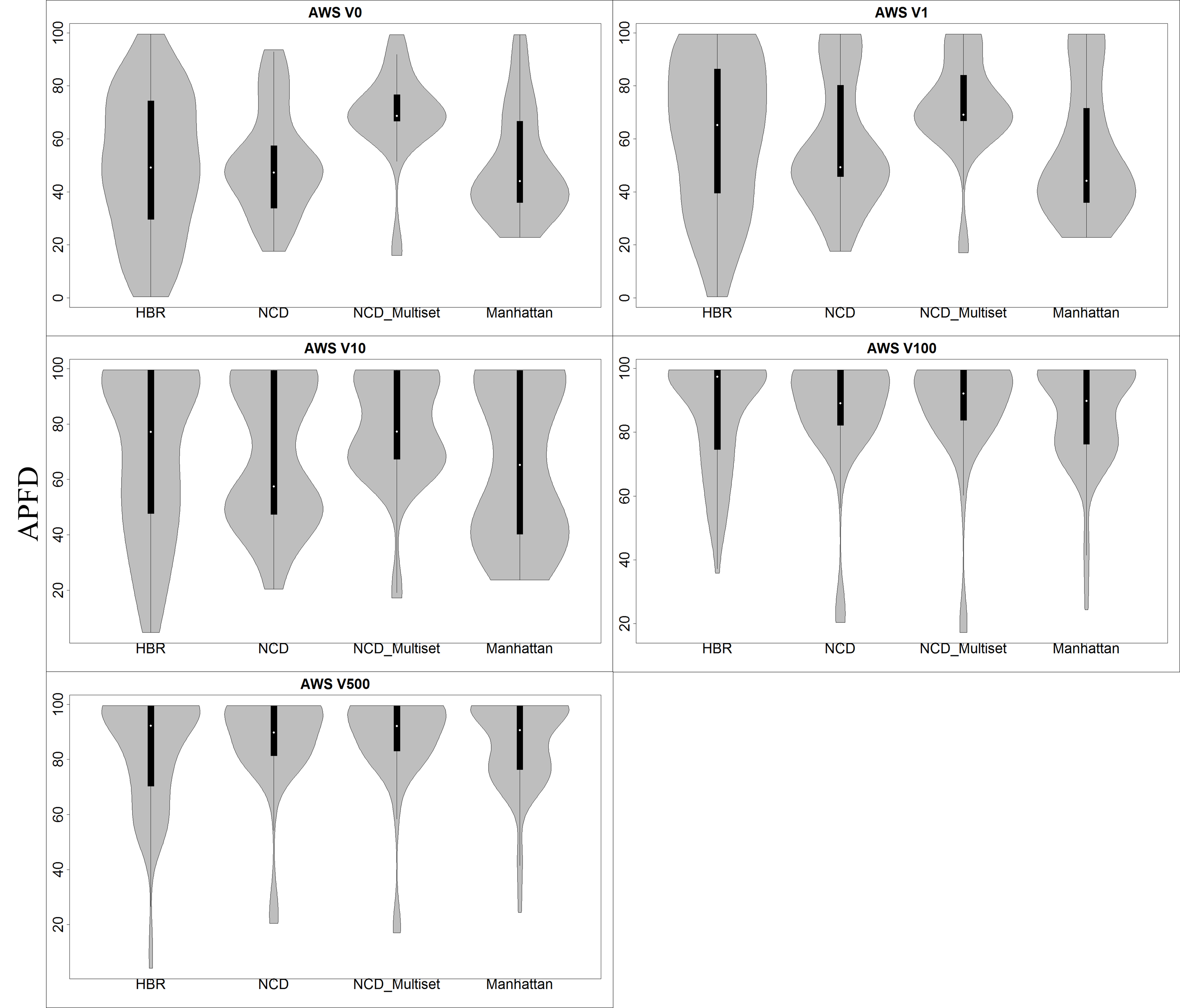}
\caption{AWS Java SDK\textendash HBD vs. HBR over different interval sizes (V)} \label{fig:RQ3_3}
\end{figure}

\begin{figure}[H]
\centering
\includegraphics[width=1\columnwidth]{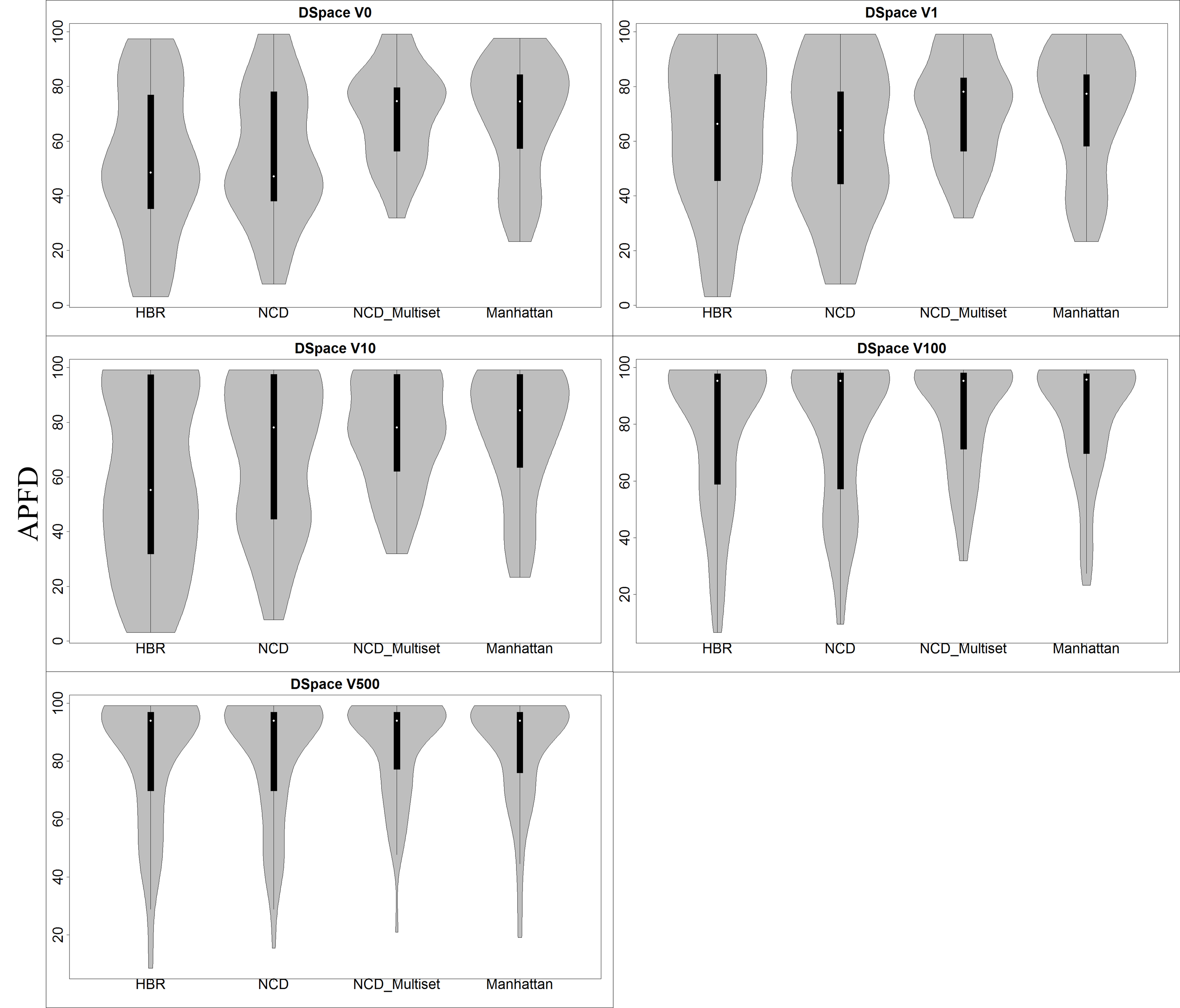}
\caption{DSpace\textendash HBD vs. HBR over different interval sizes (V)} 
\label{fig:RQ3_4}
\end{figure}

\begin{figure}[H]
\centering
\includegraphics[width=1\columnwidth]{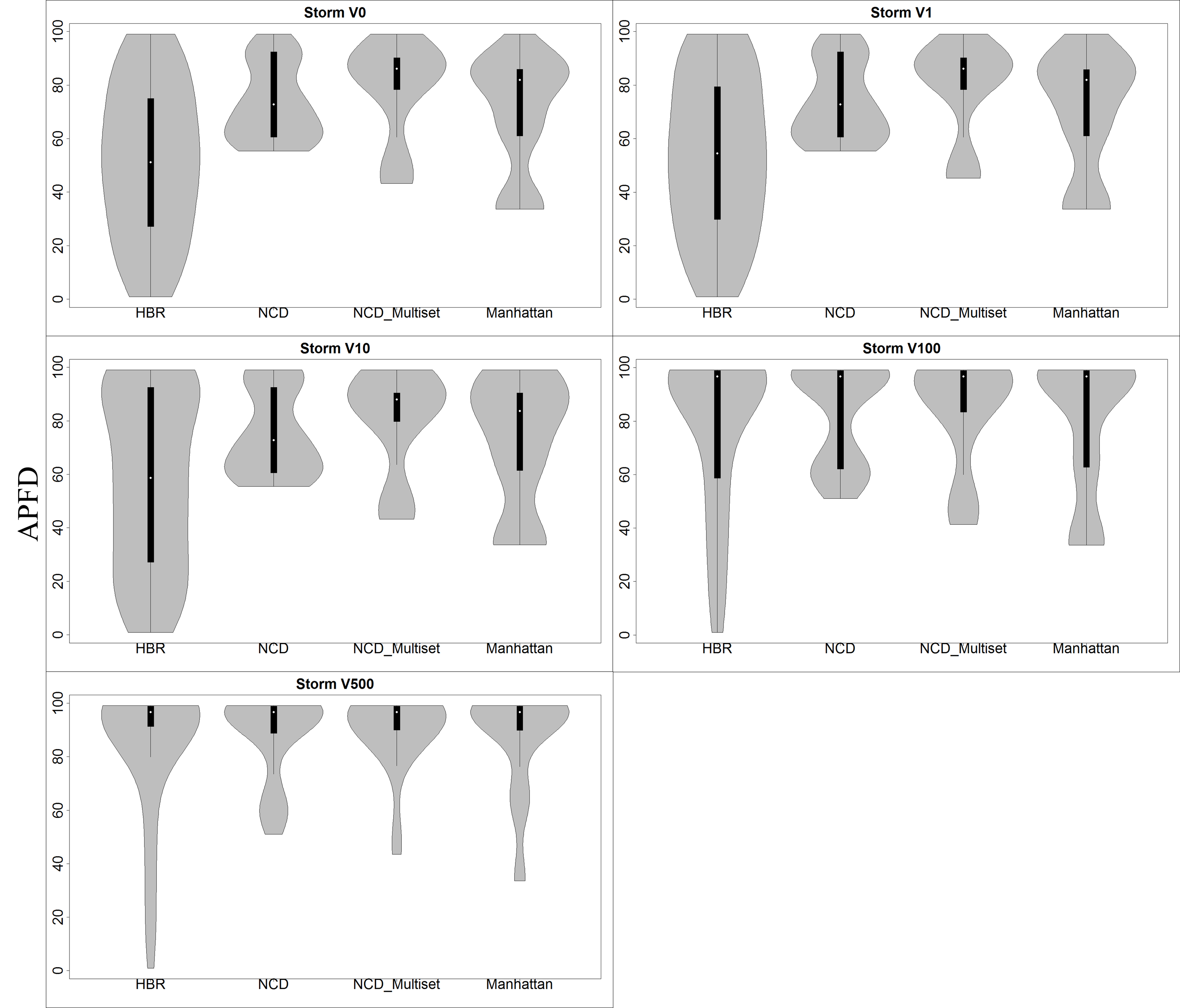}
\caption{Apache Storm\textendash HBD vs. HBR over different interval sizes (V)} \label{fig:RQ3_5}
\end{figure}

\begin{figure}[H]
\centering
\includegraphics[width=1\columnwidth]{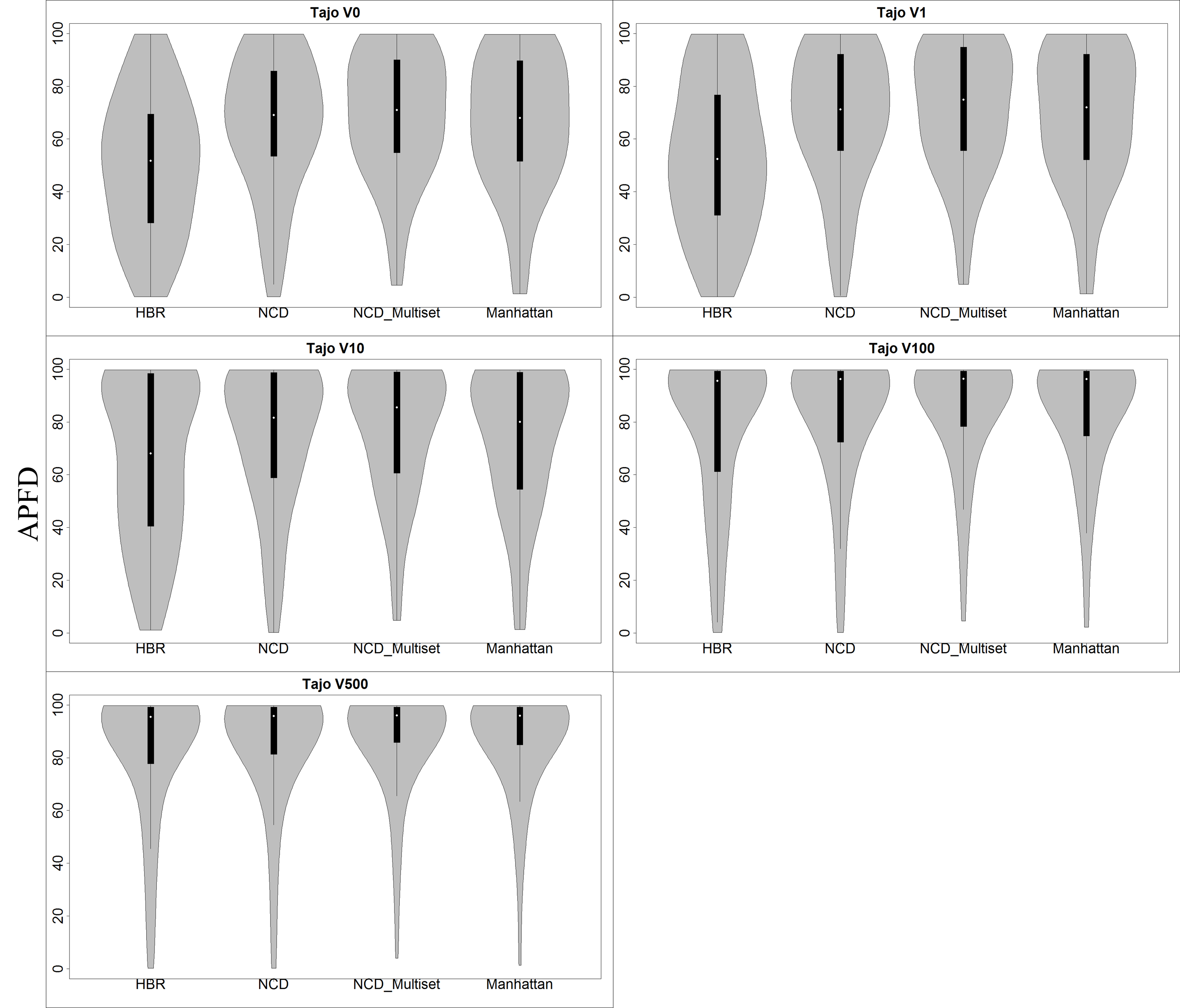}
\caption{Apache Tajo\textendash HBD vs. HBR over different interval sizes (V)} \label{fig:RQ3_6}
\end{figure}

\end{document}